# Proposition pour une gestion dynamique de l'inter-activités dans le TCAO

*Grégory Bourguin*

LIL
50, rue Ferdinand Buisson - BP 719
62228 CALAIS Cedex, France
bourguin@lil.univ-littoral.fr

**RESUME**
Nous fondant sur des résultats empiriques et théoriques provenant des sciences humaines et sociales, nous travaillons sur le problème encore prégnant de la malléabilité au sein des environnements de TCAO. Notre proposition veut favoriser l'intégration dynamique de collecticiels dans un environnement global et intégré qui crée un contexte pour leur utilisation. Ces travaux nous ont amené à poser le problème de la gestion de l'inter-activités. Cette étude nous a permis, dans une première étape, de proposer une solution technique aux problèmes posés et qui est mise en œuvre dans la plateforme CooLDA.

**MOTS CLES :** TCAO, Collecticiels, malléabilité, inter-activités, intégration.

**ABSTRACT**
Using some results coming from human and social sciences, we are working on the still important problem of tailorability inside CSCW systems. Our proposition aims at favouring the dynamic integration of groupware systems in a global and integrated environment that creates a context for their use. This work leads us to define the problem of the inter-activities management. This study helps us to propose a technical solution to this problem and that is realized in the CooLDA platform.

**CATEGORIES AND SUBJECT DESCRIPTORS:** H.4.1 [Office Automation]: Groupware

## INTRODUCTION

Une étude récente à propos des systèmes supportant l'édition coopérative sur le Web [9] souligne que malgré la pléthore d'outils existants pour supporter cette activité, aucun ne remplis totalement le cahier des charges qui semble à priori le mieux répondre aux besoins des utilisateurs. Ces remarques sont caractéristiques d'une problématique majeure dans le domaine de recherche sur le Travail Coopératif Assisté par Ordinateur (TCAO) et nous renvoient directement vers les travaux et résultats fournis ces dernières années par les Sciences Humaines et Sociales (SHS). Ces travaux nous laissent à penser que l'outil de TCAO ou collecticiel complet n'existe pas. Les théories sur l'action située [8] ou encore la Théorie de l'Activité [2] nous montrent qu'il semble impossible de connaître à priori l'ensemble des besoins des utilisateurs envers leur système. Ces besoins émergent au cours de l'activité supportée par le système, lors de son utilisation.

C'est pourquoi nous travaillons depuis plusieurs années sur le principe de malléabilité et les moyens qui permettent de la réaliser. À partir de fondements sur l'activité humaine, nous pensons qu'une application de TCAO malléable doit permettre à ses utilisateurs d'intégrer dynamiquement les outils dont ils ont besoin. C'est dans ce cadre que nous avons défini le principe de Co-évolution [3]. Nos travaux se réalisent aujourd'hui dans la plateforme CooLDA (Cooperative Layer supporting Distributed Activities). CooLDA veut fournir un environnement global et intégré de TCAO. Notre but est de permettre à un groupe d'utiliser et de faire évoluer un environnement qui crée un contexte pour l'utilisation de divers collecticiels.

Cet enjeu nous amène à étudier dans une première partie les problèmes liés au besoin crucial et souvent mal pris en compte de la gestion de l'inter-activités. La seconde partie de ce papier présente l'approche que nous développons dans CooLDA et une solution technique pour répondre au problème de malléabilité par l'intégration dynamique d'outils disponibles sur le Web.

**LE PROBLEME DE L'INTER-ACTIVITES**
Nous avons souligné le fait qu'il semble impossible de créer un environnement de TCAO qui satisfasse à priori complètement les besoins de ses utilisateurs. La solution proposée par notre approche dans CooLDA est de supporter et de favoriser l'extension de l'environnement de TCAO par la Co-évolution [3] en permettant l'intégration dynamique de divers collecticiels répondants chacun partiellement aux besoins des utilisateurs. De notre point de vue, chaque collecticiel supporte une activité composante de l'activité globale des utilisateurs. Par exemple, une messagerie instantanée peut supporter l'activité de discussion synchrone d'une activité de brainstorming. CooLDA s'intéresse donc à l'intégration et l'articulation de ces diverses activités. C'est pourquoi nous parlons du problème de la gestion de l'*inter-activités*.

**Un besoin d'intégration fine et dynamique**
Le principe d'intégration d'activités est synthétisé dans la Figure 1.

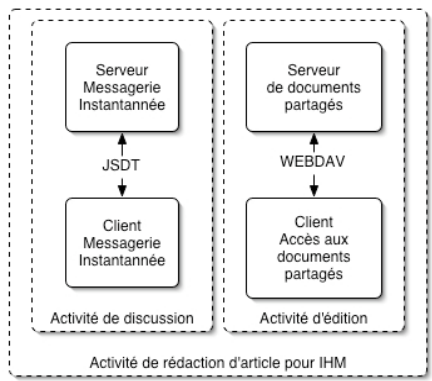

*Figure 1 : Un exemple de contextualisation d'activités*

Dans cet exemple, le serveur et le(s) client(s) de messagerie instantanée fondés sur JSDT [5] existent dans l'Internet. Ils supportent une activité de discussion synchrone. De la même manière, une application Web fondée sur le protocole WEBDAV [10] supporte une activité de partage de documents asynchrone. L'objectif de CooLDA est de fournir le support à l'activité de rédaction d'article qui contextualise les activités de discussion et de partage de documents. La contextualisation d'une application suppose que CooLDA puisse la démarrer, la piloter et, dans le cas d'une intégration très fine, recevoir des évènements concernant les changements d'états de l'activité qu'elle supporte. Le démarrage de l'application semble un besoin évident : un utilisateur qui se connecte à l'environnement CooLDA et entre dans l'activité de rédaction d'article doit se voir proposer les clients de messagerie instantanée et de partage de documents. Dans le cas d'une intégration plus fine, le rôle d'un utilisateur dans une activité peut créer une incidence sur son rôle dans les activités liées. Prenons cette fois en exemple une activité d'enseignement à distance. Dans ce contexte, le rôle de *professeur* dans un cours lui donnera le rôle de *présentateur* dans l'activité supportée par un outil WYSIWIS (What You See Is What I See) de partage de documents, et d'*orateur* dans l'audioconférence associée. Pour ce faire, l'environnement qui instancie le cours doit pouvoir piloter ou paramétrer les outils impliqués. Enfin, dans un cas d'intégration maximale, l'action d'un utilisateur jouant un rôle particulier dans une activité peut aussi avoir une influence sur l'état des autres activités. Par exemple, on peut imaginer une activité de débats mettant en œuvre un outil de vote et un forum. La proposition d'une motion dans l'outil de vote peut générer un changement de phase dans l'activité de débats et clore les discussions dans l'activité supportée par le forum associé. L'environnement CooLDA doit donc pouvoir recevoir des évènements générés par un outil et piloter les activités liées, i.e. déclencher les opérations permettant un changement d'état dans les outils qui les supportent.

Le paragraphe précédent présente les mécanismes nécessaires au support des liens tissés entre un outil externe et CooLDA. Dans notre approche de la malléabilité, il faut que notre plateforme fournisse aussi les moyens de créer et faire évoluer dynamiquement ces liens. L'utilisateur doit pouvoir découvrir dynamiquement les outils ainsi que leurs fonctionnalités (ou méthodes dans une approche objet). Nous n'aborderons pas ici les problèmes posés par cette approche du point de vue de la sémantique et de l'utilisabilité, même si ces dimensions sont aussi cruciales. En effet, la mise en œuvre dynamique de fonctionnalités implique leur découverte, mais aussi leur compréhension par l'utilisateur. Nous avons déjà commencé des travaux sur ces problèmes et proposé un embryon de solution mettant en œuvre un méta modèle inspiré de la TA dans [4]. Notre objectif est ici différent. Il s'agit dans d'identifier une plateforme qui fournisse les moyens techniques de l'intégration dynamique.

**Un environnement d'intégration dynamique simple**
L'intégration d'outils existants suppose au minimum que leur partie cliente soit mise à la disposition des utilisateurs. Il est donc nécessaire de choisir un environnement qui facilite à la fois l'accès au outils du point de vue de l'utilisateur, mais aussi qui facilite l'apparition de ces outils. En effet, il est important de rappeler que notre objectif n'est pas de créer les outils à intégrer nous mêmes. Il est donc nécessaire que nous tentions aussi de limiter les contraintes imposées par notre approche du point de vues des développeurs d'outils de TCAO comme des 'chats' ou des éditeurs partagés. Il est à noter que les mécanismes requis pour CooLDA existent dans divers environnements distribués tels que Corba. Des mécanismes similaires sont aussi disponibles dans l'approche plus récente des Services Web [7][11]. Le problème de telles approches est qu'elles requièrent la mise en place de l'environnement qui y est associé. La mise en œuvre de Services Web dans une plateforme comme J2EE nécessite par exemple l'utilisation d'un dépôt de composant particulier comme Tomcat dans lequel les services doivent être enregistrés grâce à des fichiers de définitions supplémentaires écrits en XML. Cette approche impose au développeur des contraintes fortes et un surcroît de développement dans le seul objectif d'une hypothétique intégration de son outil dans un environnement tiers comme CooLDA. C'est pourquoi nous avons choisi de ne pas mettre en œuvre ces technologies. Néanmoins, le Web semble a priori une bonne solution pour la publication et la découverte d'applications de TCAO à intégrer. Cette plateforme a été pensée pour permettre aux personnes d'échanger de l'information. Le succès du Web est en grande partie dû au fait que les navigateurs fournissent un moyen multi plateforme pour accéder à l'information, le protocole HTTP est le standard pour son transport et la pléthore de serveurs Web offre un moyen simple et peu coûteux pour la publier.

### LA PLATEFORME COOLDA

Ces divers éléments nous ont amené à une solution simple : l'environnement d'intégration utilisé par CooLDA est la machine virtuelle Java. Java a été pensée pour résoudre le problème de l'hétérogénéité. Ceci permet d'imaginer des programmes pouvant aisément être exécutés sur les plateformes les plus répandues. Il est à noter que Java est souvent utilisé comme palliatif dans les applications Web de TCAO : des environnements très connus comme BSCW [1] l'utilisent pour faciliter l'ajout de dimensions synchrones à leurs activités. L'existence dans le langage Java des nombreux frameworks tels que Corba ou RMI (Remote Method Invocation), qui supportent les systèmes à objets distribués, JSDT (Java Shared Data Toolkit) [5] dédié au TCAO, ou même le jeune JXTA (JuXTApose) [6] pour mettre en oeuvre des mécanismes Pair à Pair (P2P), facilite grandement le développement de collecticiels. Une autre particularité de Java est qu'il facilite la mise en œuvre de code mobile. Java permet aisément de télécharger des programmes au travers du réseau pour les instancier dans la machine virtuelle Java située sur le client : il est possible de télécharger et instancier dynamiquement toute application Java dont les classes ont été rendues disponibles sur Internet en ajoutant les fichiers ou archives correspondantes dans un simple serveur Web.

### Support de l'inter-activités

Une fois l'objet-application instancié dans CooLDA, nous profitons des mécanismes d'envois de messages existants dans Java pour supporter les interactions entre les activités. Les mécanismes mis en œuvre sont synthétisés dans la Figure 2. Cette figure reprend l'exemple de l'activité de débats que nous avons proposée ci avant. L'action (1) d'un utilisateur dans son client de vote en proposant une motion prévient (2) le serveur de vote, ce qui correspond au fonctionnement « normal » de cette application. Le client CooLDA se trouvant dans la même machine virtuelle peut recevoir un message/événement (3) généré par le client de vote. Ce message est remonté (4) au niveau du serveur d'activités CooLDA (par Corba dans la version actuelle) qui lui donne un sens dans l'activité globale de débats. Ceci génère un appel de méthode (5) sur le client CooLDA qui va lui-même piloter les sous activités, c'est-à-dire déclencher (6) si besoin des méthodes sur les applications qui se trouvent sur le poste client. Dans notre exemple, le client CooLDA demandera au client du Forum de faire passer son activité (par envoi de message à son serveur par exemple) dans un état où les discussions sont closes.

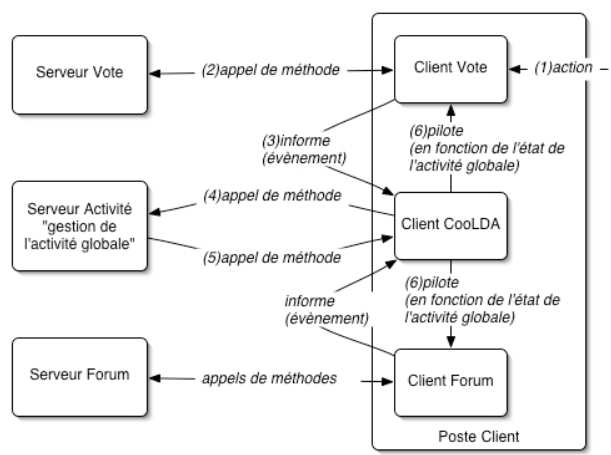

*Figure 2 : un exemple de communication inter-activités*

Il est important de noter que chaque outil est libre de mettre en œuvre son propre protocole de communication (Corba, RMI, JSDT, JXTA, etc.) pour la gestion de sa propre activité. Par contre, pour une intégration fine il est nécessaire que le concepteur de l'application ait prévu des méthodes au niveau du client permettant la communication avec un environnement tiers comme CooLDA. Dans notre exemple, il faut entre autres que le client du Forum propose une méthode du type *stopperDiscussion()* qui, lorsqu'elle est déclenchée, va changer l'état de cette activité au niveau du serveur de Forum. Ce serveur informera lui-même les autres clients du Forum sur le réseau. Toute approche voulant faire interagir des applications nécessite la création d'une telle interface. Néanmoins, notre approche a pour bienfait de ne pas solliciter un protocole de communication particulier pour les échanges entre les différentes activités. Seul le mécanisme d'envoi de message disponible dans toute machine virtuelle Java est sollicité. De plus, une partie des méthodes de cette interface existe certainement déjà au niveau du client. En effet, il est probable que le développeur de l'application ait créé un menu ou un bouton qui fait déjà appel à cette méthode qui est du niveau d'abstraction de l'activité de Forum. La dimension la plus problématique reste celle de la récupération des évènements (étape *(3)* dans la figure). Il est nécessaire que le concepteur d'application fournisse la méthode permettant à CooLDA de s'y abonner et qu'il spécifie les types d'évènements qui peuvent être générés. Ce problème n'est pas spécifique à notre approche mais est commun à toute démarche pour l'intégration dynamique et fine de composants. La solution technique que nous proposons ici supporte ces mécanismes mais pose des problèmes de sémantique que nous n'avons pas la place de développer dans ce papier.

**Spécification dynamique de l'inter-activités**

Pour supporter la spécification dynamique de l'inter-activité, nous mettons en œuvre les mécanismes réflexifs de Java, en particulier l'introspection. En effet, une fois la classe d'un objet-application Java téléchargée du site Web qui l'héberge, avant ou après son instanciation, il est possible d'obtenir dynamiquement la liste de ses méthodes. Ces mécanismes sont similaires à ceux qui sont mis en place dans la BeanBox de Sun et qui permet de lier graphiquement des JavaBeans. La différence dans CooLDA est que nous utilisons ces mécanismes pour créer des liens entre des composants de haut niveau d'abstraction. Une fois encore, il est important de souligner que cette approche permet de découvrir l'interface d'intégration et de créer des liens entre des applications sans solliciter des moyens supplémentaires. Le développeur de la partie cliente de l'outil intégrable ne doit pas générer et enregistrer une interface particulière dans un dépôt de composants. Il suffit que cette partie cliente soit écrite en Java. Néanmoins, nous souhaitons à l'avenir nous inspirer plus avant de l'approche JavaBeans et fournir quelques recommandations simples aux développeurs pour faciliter le travail d'intégration. De la même manière que la recommandation JavaBeans permet de filtrer les méthodes d'accès aux propriétés d'un objet, nous souhaitons spécifier des conventions de nom permettant de filtrer l'interface des outils intégrables. Le but est de faciliter le processus d'intégration d'un outil en ne présentant à l'utilisateur que les méthodes utiles dans ce processus. Il s'agira alors d'une recommandation du type « TCAOBeans » ou « ActivityBeans ». Le nom et la recommandation restent à spécifier.

**CONCLUSION**

De manière à répondre aux besoins émergents des utilisateurs, CooLDA se veut un environnement global et intégré de TCAO, favorisant l'intégration dynamique et l'articulation de collecticiels disponibles sur le Web. Nous avons présenté les problèmes liés à la gestion de l'inter-activités et proposé une solution simple. Cette solution permet l'intégration dynamique et l'articulation d'outils dont la partie cliente est disponible sur le Web. Elle ne requiert que le langage Java du côté client et de simples serveurs HTTP. Elle limite ainsi les contraintes liées à la mise en place d'un environnement d'exécution et qui sont posées aux futurs utilisateurs de CooLDA et développeurs d'outils ou collecticiels intégrables dans une plateforme tierce comme la nôtre. Toutefois, ces travaux doivent être développés plus avant. CooLDA fournit aujourd'hui une base solide pour explorer et expérimenter des solutions au problème récurrent de l'utilisabilité dans les environnements de TCAO malléables. Ces considérations d'un niveau plus sémantique constituent sans aucun doute une prochaine étape incontournable dans le développement de notre approche de la malléabilité par la Co-évolution [3].